\documentclass[useAMS,usenatbib]{mn2e}

\usepackage[english]{babel}
\usepackage{amssymb}
\usepackage{rotfloat}
\usepackage{pbox}
\usepackage{booktabs}

\usepackage{enumerate}
\usepackage{epsfig,amsmath}
\usepackage{graphics}
\usepackage{boldtensors}
\usepackage{color}

\DeclareMathSymbol{\Natural}{\mathbin}{AMSb}{"4E}
\DeclareMathSymbol{\Integer}{\mathbin}{AMSb}{"5A}
\DeclareMathSymbol{\Real}{\mathbin}{AMSb}{"52}
\DeclareMathSymbol{\Rational}{\mathbin}{AMSb}{"51}
\DeclareMathSymbol{\Imaginary}{\mathbin}{AMSb}{"49}
\DeclareMathSymbol{\Complex}{\mathbin}{AMSb}{"43}

\def\var{{\rm var}}
\def\Pr{{\rm Pr}}

\title[Significance of periodicities]{A comparative study of four significance measures for periodicity detection in astronomical surveys}
\author[M. S\"{u}veges et al.]{Maria S\"{u}veges$^1$\thanks{E-mail: Maria.Suveges@unige.ch}, Leanne P. Guy$^1$, Laurent Eyer$^2$,  \and
Jan Cuypers$^{3}$, Berry Holl$^{1,2}$, Isabelle Lecoeur-Ta\"ibi$^1$, Nami Mowlavi$^{1,2}$,  \and Krzysztof Nienartowicz$^1$, Diego Ord\'o\~nez Blanco$^1$, Lorenzo Rimoldini$^1$, \and Idoia Ruiz$^1$ \\
$^1$ Department of Astronomy, University of Geneva, Chemin d'Ecogia 16, CH-1290 Versoix, Switzerland \\ 
$^2$ Department of Astronomy, University of Geneva,  Chemin des Maillettes 51, CH-1290 Sauverny,  Switzerland \\
$^3$ Royal Observatory of Belgium, Ringlaan 3, 1180 Brussel, Belgium
}

\begin{document}

\date{}

\pagerange{\pageref{firstpage}--\pageref{lastpage}} \pubyear{2015}

\maketitle

\label{firstpage}

\begin{abstract}
We study the problem of periodicity detection in massive data sets of photometric or radial velocity time series, as presented by ESA's Gaia mission. Periodicity detection hinges on the estimation of the false alarm probability (FAP) of the extremum of the periodogram of the time series. 
We consider the problem of its estimation with two main issues in mind. First, for a given number of observations and signal-to-noise ratio, the rate of correct periodicity detections should be constant for all realized cadences of observations regardless of the observational time patterns, in order to avoid sky biases that are difficult to assess. Second, the computational loads should be kept feasible even for millions of time series. Using the Gaia case, we compare the $F^M$ method \citep{paltani04, schwarzenberg-czerny12}, the Baluev method \citep{baluev08} and the GEV method \citep{suveges14}, as well as a method for the direct estimation of a threshold. Three methods involve some unknown parameters, which are obtained by fitting a regression-type predictive model using easily obtainable covariates derived from observational time series. We conclude that the GEV and the Baluev methods both provide good solutions to the issues posed by a large-scale processing. The first of these yields the best scientific quality at the price of some moderately costly pre-processing. When this pre-processing is impossible for some reason (e.g. the computational costs are prohibitive or good regression models cannot be constructed), the Baluev method provides a computationally inexpensive alternative with slight biases in regions where time samplings exhibit strong aliases.
\end{abstract}


\section{Introduction} \label{sec_intro}

Identification and analysis of variable objects has been and will remain an important product of many small or large-scale astronomical surveys. Periodic sources among them are singularly important for many special fields of astrophysics. Examples include the Cepheids, which form the basis of the cosmic distance ladder, and therefore are fundamental to cosmology; RR Lyrae, which trace ancient structures around the Milky Way and thus relate to the evolution of our Galaxy; multiperiodic stars, whose asteroseismology provides insight into the structure and evolution of stars; or eclipsing binaries, which can offer the possibility of determining the mass and radius of their component stars, and thus also provide strong constraints on stellar physics and the co-evolution of stars.

Specific research fields require the selection of objects of specific types, and the databases in which they must be sought  have now reached the terabyte scale. Whereas the Hipparcos mission \citep{hipparcos} twenty years ago provided photometry and astrometry for about a hundred thousand stars, the Gaia mission, launched in December 2013, will furnish a similar  catalog of approximately 1 billion astronomical objects by the end of its 5 year lifetime, with a precision of roughly 100 times that of Hipparcos. Other surveys also will produce databases of comparable or larger size. To facilitate efficient searches on such volumes, the catalogs should contain additional derived information about the sources. Specifically, for studies relying on variable stars, the variability type is of primary importance, however other attributes of the source such as the mean absolute magnitude, period, amplitude, harmonic amplitudes and relative phases can all help the researcher to direct better his/her search for a sample. 

A crucial step to obtain this information is the discovery and correct identification of the period of the variable objects. A search for periodicity is performed on the time series of either photometric or radial velocity measurements of candidate objects, in order to produce a periodogram. The found period corresponds to the extremum of the periodogram.  A decision should then be made as to whether this period is significant or not. Depending on the decision, the process then can continue with the characterization of the source as periodic and the production of the derived information for the catalog. When the decision step fails for a source, this information will obviously be erroneous. If these failures are systematic, and depend on some unrecognized factors, studies using such samples may be affected by serious unidentified and unknown biases. 

Unfortunately, period detection is one of the procedures which is most at risk from such biases, because quasi-periodicities and sparse sampling in the time cadence of the observations affect the statistical characteristics of the periodogram. Their most important effects are the strong long-term dependencies appearing in the periodograms (called ``aliases" in the rest of this paper), the loss of an orthogonal frequency system (that is, loss of orthogonality of the Fourier frequencies), and the degeneracy caused by computing an oversampled periodogram often at hundreds of thousands of test frequencies, based often on only a few dozen observations. Nevertheless, whether a found period belongs to a real periodic signal or is just due to random fluctuations must be assessed in a strictly formalised statistical way \citep[a concise and clear paper is ][]{schwarzenberg-czerny98}. In the presence of these strong long-range dependencies, the most commonly applied statistical tests \citep{lomb76, scargle82, hornebaliunas86, frescuraetal08, schwarzenberg-czerny12} lack solid theoretical support, and can yield incorrect estimates in the absence of clear recipes by which to tune them \citep{hornebaliunas86, frescuraetal08, schwarzenberg-czerny12}.  Thus, they can present largely uninvestigated biases. When applied en mass to  time series from a sky-scanning survey, which have coordinate-dependent observational cadences with different regularities from point to point, these biases will add an unknown, coordinate-dependent element to other, better investigated biases, such as that due to the number of observations \citep[for e.g. the Gaia survey, see][]{eyermignard05}.  

In addition to these biases, the computational greediness of most procedures makes the situation even more difficult for survey data. The above methods usually need many noise simulations for each time sampling pattern as well as the computation of the periodogram for each simulation, in order to reproduce the distribution of the periodogram peak in the absence of periodicity. This is not feasible during the data processing of a large-scale survey producing millions of time series.

To help solve these issues, we collected three propositions from the literature how to perform a significance test on periodograms: that of \citet{paltani04} and \citet{schwarzenberg-czerny12} ($F^M$ method), that of \citet{baluev08} (Baluev method) and that of \citet{suveges14} (GEV method), and we added a fourth, {\it ad hoc} one, which consists of the direct determination of a critical level of periodogram peaks separating significant periodicities from nonsignificant ones at a given confidence level (quantile method). Three of these models, the $F^M$, the GEV and the quantile methods, depend on unknown parameters, which differ from one sky location to another, and must be estimated for each of the candidate variable light curves (several tens of millions for Gaia).  

In our study, instead of individually estimating these parameters for each time series, we investigated how they depend on some quantities that can be easily and quickly computed for every time series, such as the number of observations, the variance of times or spectral window features. We constructed regression-type models linking these covariates to the parameters of the FAP methods. As a result, the costly simulation-based individual estimation of the parameters can be replaced with an estimation based on only the calculation of the above quantities and predicting the parameters from the previously estimated regression model with excellent results. 

In order to achieve this, the crucial condition is the existence of such a regression model. Although theory gives indications  as to what covariates the parameters of the FAP methods may depend on, at present there is no derivation of specific formulae or relationships. 
For the Gaia case, where the observational times consist of relatively irregularly spaced clusters of quasi-regular sequences of observations, some quite clear-cut relationships were found empirically. Since for many scanning surveys, their location on Earth or in a space orbit and/or their rotation determines some typical repeating observational cadences, and hence some characteristic spectral window patterns, in general it appears worthwhile to investigate these possibilities for the detection of periodic variability in other surveys too.

The performance of the procedures was assessed using two fundamental statistical paradigms. First, on simulated noise sequences, we checked the false alarm rate of the methods and the quality of their approximation to the true distribution of the maximum of the periodograms. Second, on weak noisy signals, we checked the ability of the methods to find their periodicity as significant. This way, we characterize the methods in terms of their statistical size and power. We conclude that at least two of these methods, the Baluev and the GEV ones, provide good approximations to the p-value of the periodicity in the interesting low value range.

In Section \ref{sec_gaia} we give an overview of the problem. First we summarize the statistical principles applied in the detection of periodicities, and discuss the factors that can influence the crucial ingredient in the methods, the distribution of the extrema of the periodograms of white noise. We demonstrate these effects on a simple model using Gaia-like simulations. Section \ref{sec_meth} presents the four candidate methods and the regression models to estimate their parameters. Section \ref{sec_app} details their application to the Gaia survey, and summarises the expected performances of the four methods using noise and signal simulations. Finally, Section \ref{sec_disc} provides a summary table of the crucial advantages and drawbacks of the methods, and discusses the possible choices for large surveys.

\section{Periodicity detection}\label{sec_gaia}

\subsection{Principles of testing}\label{subsec_princip}


Suppose $X_1, X_2, \ldots, X_N$ is a photometric or radial velocity time series observed at epochs $t_1, t_2, \ldots, t_N$. The sequence of times may be anything from almost completely irregular to almost completely regular. The goal is to assess whether the time series $X_1, X_2, \ldots, X_N$ contains a periodic signal or not. To this end, we compute a periodogram, consisting of some appropriately defined goodness-of-fit measures of some periodic models at a large set of candidate frequencies, using one or more of the var\-ious methods in the literature, such as the Deeming method \citep{deeming75}; PDM-Jurkevich \citep{jurkevich71, stellingwerf78, dupuyhoffman85}; String Length \citep{laflerkinman65, burkeetal70, renson78, dworetsky83, clarke02}; SuperSmoother \citep{friedman84, reimann94}; CLEAN \citep{ foster95}; Keplerian periodograms \citep{cumming04};  Lomb-Scargle or generalized Lomb-Scargle method (\citealt{lomb76,ferraz-mello81, scargle82, zechmeisterkurster09}); FastChi2 \citep{palmer09}; conditional entropy method \citep{grahametal13}; FAMOUS (F. Mignard, available with documentation at the website ftp://ftp.obs-nice.fr/pub/mignard/Famous).  The most probable frequency of a potential periodic signal is indicated by the extremum $z_{\rm{obs}}$ of the periodogram, which can be a maximum or minimum depending on the specific period search algorithm.

Strict statistical hypothesis testing contrasts the zero hypothesis $H_0$ of no periodicity to the alternative $H_1$ of a periodic component of any frequency in the time series. It consists of computing the probability that a time series with no periodicity produces a maximum higher than or equal to the observed maximum $z_{\rm{obs}}$ (or a minimum lower than or equal to the observed minimum). This probability is called the False Alarm Probability  (FAP). Denoting in general  the distribution of $z_{\rm{obs}}$ under $H_0$ with $G$, FAP $=1-G(z_{\rm{obs}})$ for maxima and  $\mathrm{FAP} =G(z_{\rm{obs}})$ for minima. If we find a periodogram extremum which is less likely than a pre-specified confidence limit $\alpha$, then we can state that the hypothesis of no periodicity can be rejected at the confidence level  $\alpha$. This case of a FAP $\leq \alpha$ will be termed a detection. Based on inspection of frequency search results for weak noisy signals, if the significant maximum/minimum of the periodogram is within  $\pm 10^{-3} d^{-1}$ of the correct frequency, then we will speak about a correct detection, otherwise an incorrect or false detection (we did not use the theoretically based formula for the calculation of the errors, since the inspection showed a significantly enlarged distribution of absolute differences between true and found frequencies with respect to what is expected from the formula).

To compute the FAP, we need to know the zero distribution $G$ of $z_{\rm{obs}}$ under $H_0$, that is, the distribution of the periodogram maximum in the absence of periodicity. This is determined by several factors.

\begin{itemize}
\item The periodogram type defines the distribution $F$ of any single periodogram value (in statistical terminology, the marginal distribution or margin), and has a determining role in shaping $G$. For example, for the generalized least squares (GLS) periodogram as defined in \citet{zechmeisterkurster09}, $z \in [0,1]$, and $F$ is approximately a beta distribution, so $G$ also must have a finite tail with endpoint at 1. The Lomb-Scargle periodogram with the original normalisation \citep{lomb76, scargle82} has an exponential marginal distribution, with an exponentially decaying tail, and thus $G$ must have a tail smoothly decreasing towards infinity.
\item The ``no periodicity" assumption is usually not sufficient to constrain $G$, we must make further assumptions about the character of the time series under $H_0$. Some options are: the time series is white noise with a specific distribution; the time series is white noise, with an unspecified distribution; the time series has some specified temporal structure such as a CARMA process or a random walk. The derivation of $G$ is very hard under most of these assumptions, so there is practically no known $G$ for the periodograms listed above under any assumption other than white noise, and even under the assumption of white noise, well-behaved approximate distributions were published only quite recently and only for certain types of periodograms.
\item An irregular character of the time sampling entails the loss of any orthogonal frequency system, so in theory, no principles relying on independence could be applied without an extra step of orthogonalisation. Quasi-regularities in the time sampling, e.g. daily and yearly cadences in ground-based observations or the 6-hour spin rate of Gaia, can induce aliasing (strictly speaking, leakage, but we will use the commonly accepted term in astronomy), which leads to a complex pattern of dependence across frequencies in the periodogram. Moreover, dependence is also introduced into the periodogram by the simple fact that based on $N$ observations, we usually compute $n \gg N$ periodogram values. According to mathematical statistics, extrema of dependent sets do not behave the same way as those of independent sets \citep[see e.g.][]{leadbetter, statofext, dehaanferreira}, not even when their marginal distribution is the same, so this dependence must have an effect on $G$. This is clearly demonstrated by the simulations of \citet{cuypers12}, using Gaussian noise sequences with the same time span and the same number of observations, but with different temporal sampling patterns.
\end{itemize}

Theory and simulations thus both suggest to take dependence into account when trying to derive the distribution of periodogram maxima. However, there is no general mathematical derivation pointing to explicit formulae with which this could be accomplished. The parameters of the distribution of maxima of dependent or independent variables are in practice not derived from theory, but estimated case-wise for every time series. The reason is that dependence itself in the set of random variables can take infinitely many forms, and for any real-life data set, usually little is known about it a priori.

\begin{figure}
\begin{center}
\includegraphics[scale=.3]{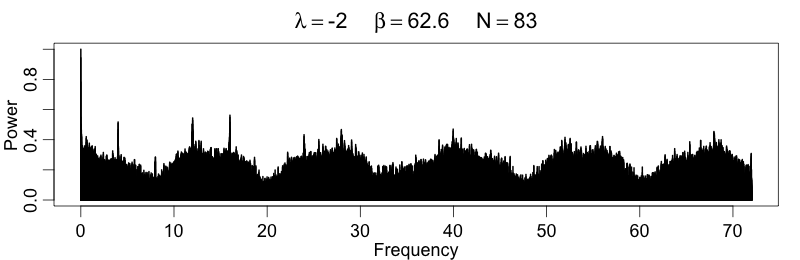}
\includegraphics[scale=.3]{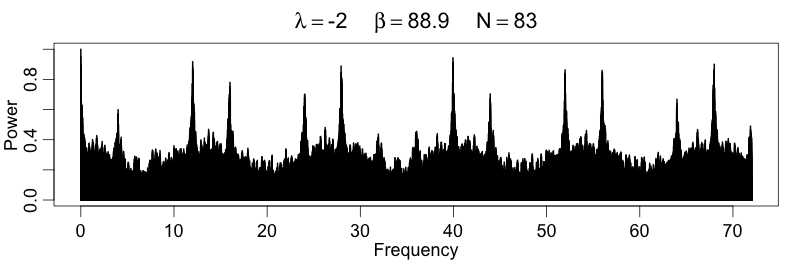}
\caption{Spectral windows of Gaia at two different ecliptic latitudes with the same number of observation.}
\label{fig:spwins}       
\end{center}
\end{figure}

\begin{figure}
\begin{center}
\includegraphics[scale=.4]{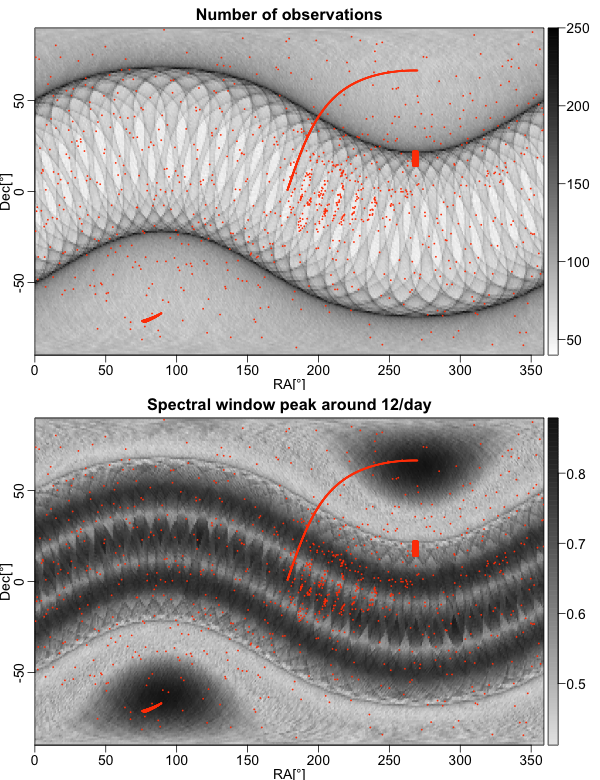}
\caption{Number of observations $N$ (top) and spectral window peak $s_{12}$ around 12 $d^{-1}$ (bottom) over the sky; the shades of grey indicate the value of $N$ and $s_{12}$. The red dots show the locations used for training, model selection and testing.}
\label{fig:map_n_at12}       
\end{center}
\end{figure}

However, for the frequency analysis of time series, we have an aid in revealing dependence structures in the periodograms: the spectral window of the time cadence yields a picture of the autocorrelations in the periodogram. Though correlations are in general not sufficient measures of dependence (apart from the case of a multivariate Gaussian), a non-zero correlation is an indicator of dependence.  Moreover, from a practical point of view, the spectral window is quite easily available. As a substitute for theory-based relations between the distribution of periodogram maxima and the dependence in the periodogram, we may look for relationships linking the parameters of the distribution to numerical features of the spectral windows.

Gaia's complex motion, consisting of a 6-hour rotation and a slow precession of its axis, induces a large variety of spectral windows showing more or less prominent peaks corresponding to its spin rate $4 d^{-1}$ and the integer multiples thereof \citep[see e.g.][]{eyermignard05}. Two examples in Figure~\ref{fig:spwins} illustrate the range of possibilities. The spatial variations over the sky, together with those of the number of observations $N$, can be appreciated in Figure~\ref{fig:map_n_at12}, which shows maps of $N$ and the strength of a typical Gaia alias, at 12 $d^{-1}$. Their spatial inhomogeneity implies that we can expect also the distribution of maxima to vary over the sky. We show on an example in the next section that this is indeed so.

\subsection{Illustration}\label{subsec_illustration}

\begin{figure}
\begin{center}
\includegraphics[scale=.34]{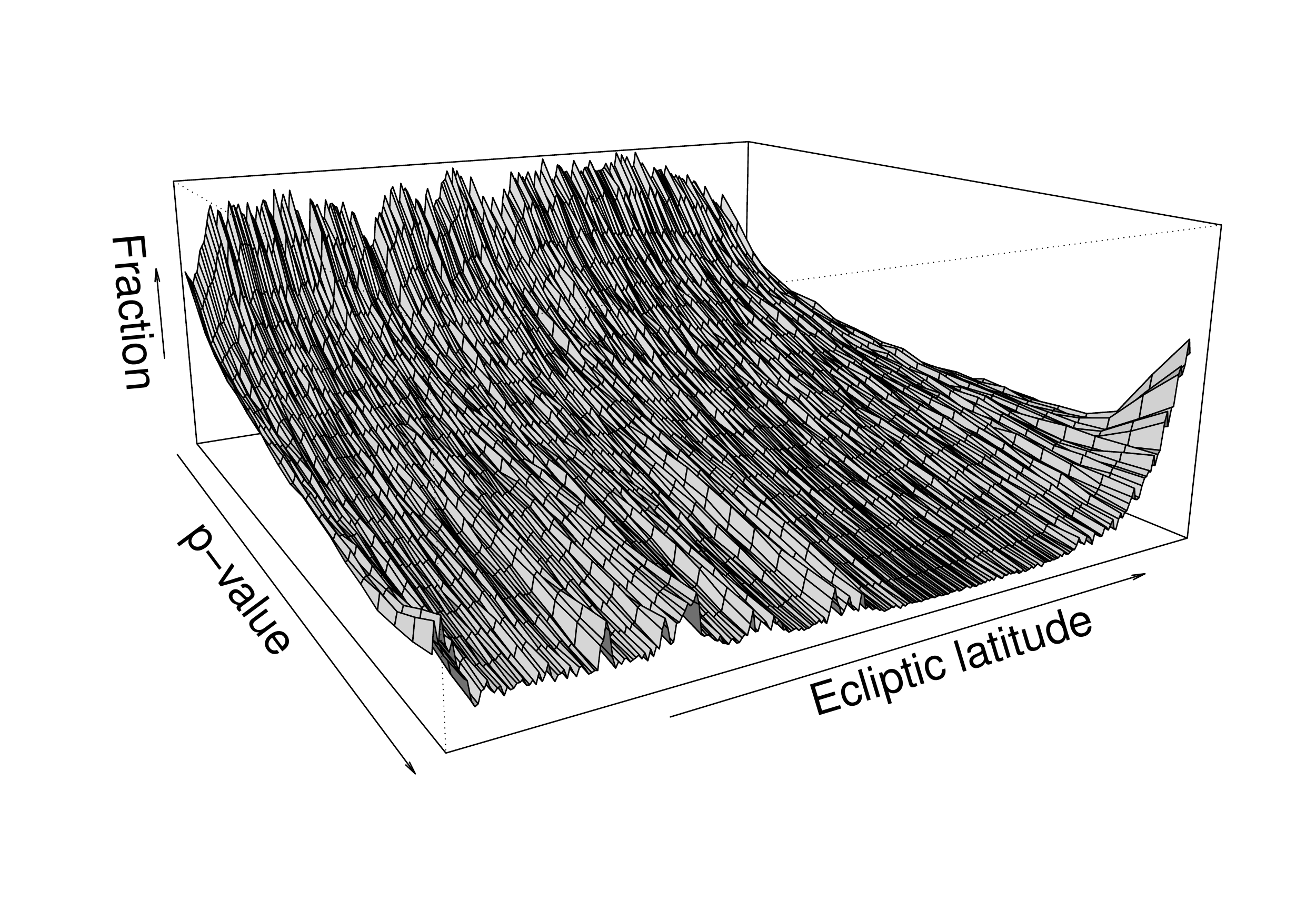}
\vspace{-1cm}

\includegraphics[scale=.68]{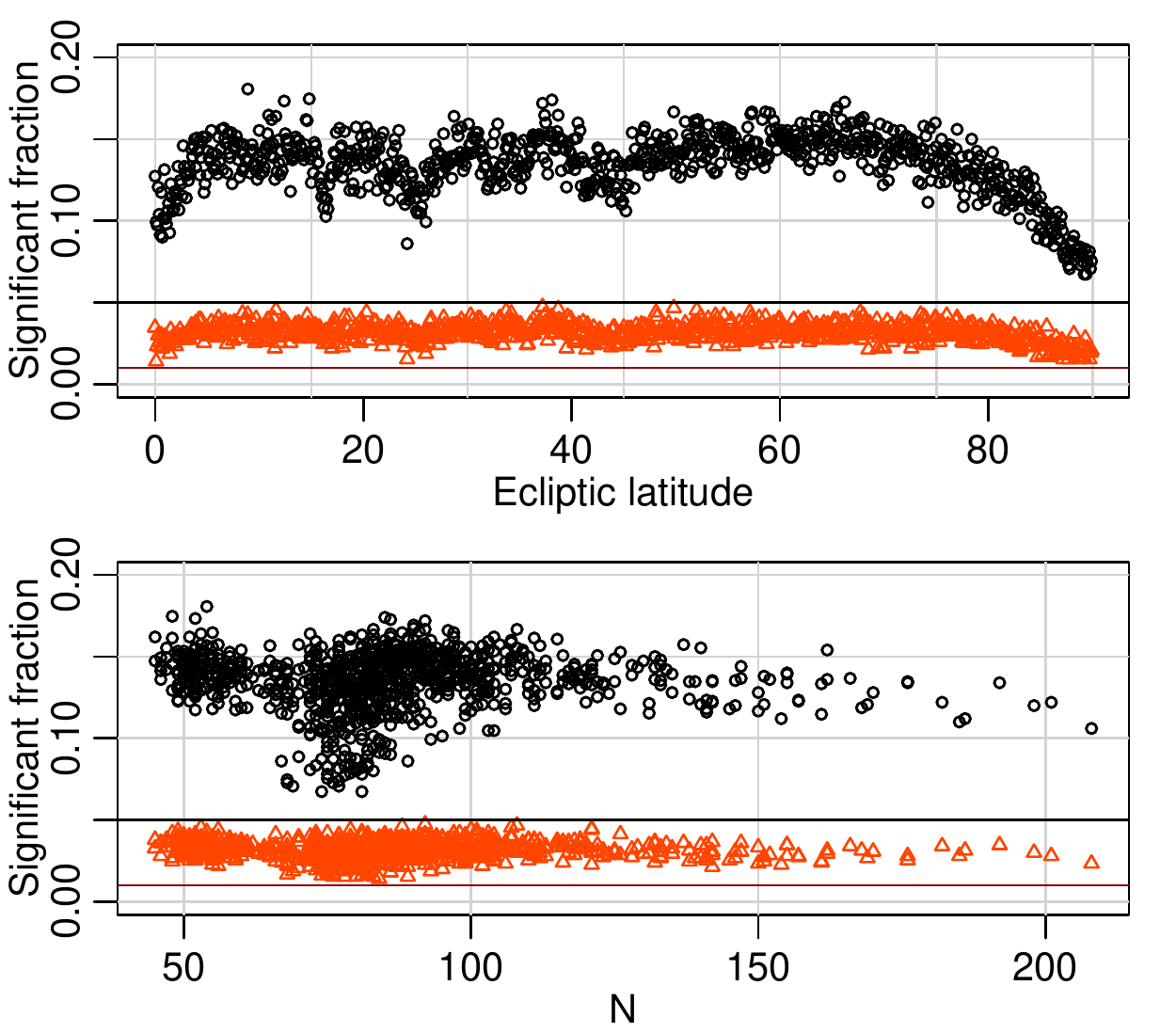}
\caption{Top panel: histograms of FAPs of white noise sequences side-by-side at 300 locations along a meridian from ecliptic to pole (top panel). Middle panel: the fractions of $p$-values less than 0.05 (black circles) and of those less than 0.01 (red triangles) versus ecliptic latitude. Bottom panel: the same versus $N$, the number of observations in the time series. The black and brown solid lines indicate the expected fractions 0.05 and 0.01, respectively.}
\label{fig:lomb_line}       
\end{center}
\end{figure}

We illustrate the importance of these issues and their possible effects on the detection of periodicities over the sky by showing the discrepancies resulting from an independence-based simple FAP model in the Gaia case. The model approximates the distribution of the periodogram maxima by 
\begin{eqnarray}\label{eq:distr_fadm}
G \approx F^M, 
\end{eqnarray}
where $M$ is the ``equivalent number of independent frequencies" \citep{scargle82, hornebaliunas86, schwarzenberg-czerny98}, which should optimally be estimated for each time sampling from a high number of noise simulations, and $F$ is the marginal distribution of the periodogram. 

We simulated 1500 Gaussian white noise sequences using Gaia-like time samplings at each of 900 sky positions along an ecliptic meridian from the ecliptic plane to the pole (the long red line in Figure~\ref{fig:map_n_at12})\footnote{See the link Gaia Observation Forecast Tool at {\tt http://www.cosmos.esa.int/web/gaia/gaia-data} for an online tool to predict Gaia observations for arbitrary sky positions.}. A generalised least squares (GLS) periodogram \citep{zechmeisterkurster09} was calculated for each sequence, its maximum retained, and the FAP based on equation \eqref{eq:distr_fadm} computed. For the purposes of the illustration, and since many simulations per time series are in any case unfeasible in large surveys, we replaced $M$ by its upper limit $M'$, the number of Fourier frequencies falling in the scanned frequency range. The marginal distribution of the periodogram values is approximately a beta distribution for the GLS periodogram, so $F$ is taken to be the incomplete beta function with parameters 1 and $(N-3)/2$ \citep{schwarzenberg-czerny98}. At each location, we obtained thus 1500 FAP values for white noise sequences.

The upper panel of Figure~\ref{fig:lomb_line} shows 300 of the 900 histograms of FAPs, lined up side-by-side versus ecliptic latitude. Good approximations to the true distribution of the maxima should give a flat aspect of the surface traced by the histograms: the FAP values for noise based on a good approximate $G$ should follow a uniform distribution, that is, a flat histogram when binned, and moreover, they should do so at every location, independently of latitude or any other parameter. The lower two panels show only the last bin of these histograms, that is, the fraction of FAPs below the significance levels $\alpha = 0.05$ (black) and $\alpha = 0.01$ (red). This is the proportion of false detections (noise sequences for which a periodicity was detected). In the case of a well-behaved FAP, this should be close to $\alpha$ by definition of the FAP. From the plots, we can draw several important conclusions.
\begin{itemize}
\item The false detection fractions are in general much above $\alpha$, which means that this FAP approximation is too permissive. The best $M$ should indeed be lower than the number of Fourier frequencies $M'$.
\item The middle panel of Figure~\ref{fig:lomb_line} shows a strong spatial inhomogeneity: some regions, most importantly the ecliptic pole  ($\mid \! \beta \! \mid\; > 80^{\circ}$), have a much lower average false alarm rate than others, say, a region at $60^{\circ} < \;\mid \! \beta \! \mid\; < 70^{\circ}$.  
\item The false alarm rates depend only very weakly (if at all) on $N$, as it can be seen in the bottom panel of Figure~\ref{fig:lomb_line}, so this spatial variation is not governed by variations in $N$. Around $N=70-80$, the significant fraction varies in a much wider interval (between $0.06, 0.16$) than at other $N$ values; the lowest false alarm rates here are due to the ecliptic poles, where the numbers of observations are about 80. As the ecliptic pole is the region where aliasing is the strongest, indicating the strongest dependences in periodograms, this hints at the significant effect of dependences in the distribution of periodogram maxima.
\end{itemize}
A few important practical consequences can immediately be seen. First, two variable objects with the same light curve characteristics and with the same number of observations and signal-to-noise ratio can have different detection probabilities, if located in different sky regions. Second, inhomogeneities are present on very short angular distance scales on the sky, because the shapes of the histograms are sharply fluctuating on very short distance scales. We need to account for these spatial inhomogeneities on the sky. The plots in this illustrative example and theory both suggest that this should be pursued through a modelling of the distribution $G$ of the maxima with the help of $N$ and variables characterizing the dependence in the periodogram. In the next section, we present several different approximations to $G$, and describe the strategy to model its spatial variations.

\section{Methodology}\label{sec_meth}

\subsection{The four approximations to the FAP}\label{subsec_fapmeth}
 
There are now several propositions in the literature to obtain a reliable FAP for periodicity detection. We will investigate here three of them: the $F^M$ method \citep{paltani04, schwarzenberg-czerny12}, based on the formula $G = F^M$; the Baluev method \citep{baluev08}, based on high-level excursions of stochastic processes; and the GEV method \citep{suveges14}, based on univariate extreme-value theory of random variables. To these, we add a fourth, \textit{ad hoc} alternative: we directly estimate some desired limit level (quantile) that separates periodogram extrema that are significant at the required confidence level from non-significant ones. We applied the GLS method as described in \citet{zechmeisterkurster09} to search for periods. In this normalization, the marginal distribution of the periodogram is Beta$(1,(N-3)/2)$ \citep{schwarzenberg-czerny98}, and the extremum indicating the most likely frequency of a potential signal is its maximum.

The first three alternatives all attempt to give an approximation for the distribution of the maximum of the periodogram of a white noise process as the null distribution. The fourth one, the quantile method produces only a decision whether or not the found periodicity is significant, without giving its probability under the noise assumption. The four alternatives are the following.

\begin{description}
\item{\it $F^M$ method \citep{paltani04, schwarzenberg-czerny12}}

This method approximates the true distribution $G$ of the periodogram maxima by that of the maxima of an equivalent system of $M$ independent frequencies. These frequencies usually cannot be identified with any subset of real test frequencies, as the periodogram values are more or less dependent over the whole spectrum. The approximate null distribution is
\[
G_{F^M} = F^M,
\]
where $F$ is the marginal distribution of the periodogram \citep[Beta$(1,(N-3)/2)$ for the GLS used here;][]{schwarzenberg-czerny98}, and $M$, the only unknown parameter of the distribution, is the number of equivalent independent frequencies. $M$ is estimated using simulations; the Paltani--Schwarzenberg-Czerny proposition is to compute periodograms of a sufficiently high number of noise sequences under the same time sampling, extract their maxima, and then use the median $z_{\rm med}$ of these maxima to give an estimate for $M$ as 
\[
\hat M = \frac{\log 0.5}{\log F(z_{\rm med})} .
\]
After having estimated $M$ for a specific time sampling, the p-value of an observed periodogram maximum $z_{\rm{obs}}$ can be given by
\begin{eqnarray}\label{eq:p_fadm}
p = 1 - F(z_{\rm{obs}})^{\hat M}.
\end{eqnarray}

\vspace{1mm}
\item{\it Baluev method \citep{baluev08}}\label{subsec_baluev}

This method is based on the theory of the extremes of continuous-parameter stochastic processes with beta, Fisher-Snedecor or chi-squared margins. We use here a variant with beta margins.  An upper bound to the FAP is given using approximations to the right tail of the exact distribution as
\begin{eqnarray}\label{eq:p_baluev}
p = \frac{\Gamma((N-1)/2)}{\Gamma((N-2)/2)} \sqrt{4\pi \var(t_i)} f_u \left(1- z_{\rm obs} \right)^{\frac{N-4}{2}} \sqrt{z_{\rm obs}},
\end{eqnarray}
where $\Gamma(.)$ is the gamma function,  $\var(t_i)$ is the variance of the observation times, and $f_u$ is the uppermost test frequency of the periodogram. The derivation assumes a low level of aliasing and spectral leakage, and is best at the lowest FAP values. p-values higher than 0.05 should be considered only as rough bounds on the actual p-value, but for $p \rightarrow 0$, the method provides good approximations to the actual p-values when aliases are weak. The above formula does not contain any unknown parameters, only simple quantities like the number of time series points or the variance of the observing times, which makes its implementation and application very fast and straightforward.

\vspace{1mm}
\item{\it GEV method \citep{suveges14}}\label{subsec_evd}

A simpler look at the periodogram considers it as a set of discrete random variables with a particularly strong inter-dependence. According to univariate extreme-value theory, the maximum of a set of random variables follows the generalized extreme-value distribution (GEV):
\begin{eqnarray} \label{eq:gev}
G(z) &= & {\rm Pr} \{ Z_{\max,n} \leq z  \}  \nonumber \\
         &= &\exp  \left\{ - \left( 1 + \xi \frac{z - \mu}{\sigma} \right)^{-1/\xi} \right\}, \\
  &   & \xi \in \Real, \quad \mu \in \Real, \quad \sigma > 0,  \nonumber
\end{eqnarray}
where the distribution function is defined only on $z$ such that $1+\xi(z-\mu)/\sigma > 0$ (for the GLS periodogram the endpoint of the distribution must be 1, that is, $\mu = 1 + \sigma / \xi$). This limiting distribution plays a similar general role for maxima as the Gaussian distribution for sums and averages, and this remains so for certain types of dependence. In practice, the GEV approximation works well for astronomical periodograms when its parameters are estimated individually for every time sampling, even though mathematical theory has yet to prove the dependence in astronomical periodograms to fall under the general validity condition of extreme-value limits.

The two free GEV parameters $\xi$ and $\sigma$ must be estimated for all individual time series. Once the estimates $\hat \xi$, $\hat \sigma$ and $\hat \mu =  1 + \hat \sigma / \hat \xi$ have been obtained, the p-value of an observed periodogram peak $z_{\rm{obs}}$ can be computed as 
\begin{eqnarray} \label{eq:pgev}
p = 1 - G(z_{\rm{obs}}) = 1 - \exp  \left\{ - \left( 1 + \hat \xi \frac{z_{\rm{obs}} - \hat \mu}{\hat \sigma} \right)^{-1/\hat \xi} \right\},
\end{eqnarray}
and can be compared to the pre-specified level $\alpha$.

\vspace{1mm}
\item{ \textit{Quantile method}} 

This procedure consists of estimating directly the level $z_{1-\alpha}$, which is exceeded by a maximum produced by a pure noise sequence with the pre-specified probability $\alpha$. That is, $\Pr (Z > z_{1-\alpha}) = \alpha$ if $Z$ is the maximum of the periodogram of a white noise sequence. Figure~\ref{fig:lomb_line} shows that this critical level $z_{1-\alpha}$ (the $1-\alpha$ quantile) depends on the location, and a direct estimation of $z_{1-\alpha}$ needs to be done location-wise. Once $z_{1-\alpha}$ is estimated, the question ``Is this periodicity significant or not at the level $\alpha$?" is equivalent to the question whether the computed $z_{\rm{obs}}$ is larger or smaller than $z_{1-\alpha}$. Thus, this method does not return any p-value, only gives a yes/no answer to the question of significance, and there is no quantification how unlikely $z_{\rm{obs}}$ is to come from noise.  There are no other parameters to estimate than $z_{1-\alpha}$ itself. 

\end{description}

\subsection{Parameter estimation}\label{subsec_parest}
 
Two of the approximations for $G$, namely the $F^M$ and the GEV, and the quantile methods all contain one or more quantities that should be estimated: $\xi$ and $\sigma$ for the GEV, $M$ for the $F^M$ method, and the quantile $z_{1-\alpha}$ itself in the quantile method. In principle, the best way would be to estimate them individually for each observed sequence, by simulating a large number of noise sequences with the same marginal distribution as the observations and with the same time sampling pattern. But as this involves the computation of a lot of periodograms (of the order of 5-10 in the case of the GEV method, and hundreds or thousands for the others), this case-wise estimation cannot be applied to all objects of a large survey.

The first substitute for the case-wise estimation may be interpolation, if the characteristic cadence patterns on the sky are known in advance for the survey. We can then simulate a high number of noise sequences on a sufficiently fine grid on the sky with the local predicted time sampling, estimate the local parameters of the chosen method, and interpolate them to obtain the parameters at any other point. However, Figure~\ref{fig:lomb_line} shows  extremely sharp variations as a function of ecliptic latitude, which in fact would be even more violent if all 900 histograms could have been plotted. There is also at least one important characteristic of the time series that cannot be expected to be smooth, the number of observations $N$, which is inherently discrete. The distribution of the maxima is expected to depend on it. It follows that simple interpolation in coordinates might not yield an adequate model.

Nevertheless, for surveys with a fixed prescribed scanning law like Gaia, these relationships provide a way to avoid costly case-by-case estimation during the mission at the cost of some preparatory work. Suppose that we have a representative sample of sky locations $l = 1, \ldots, L$. We can obtain the observing times {\em in advance} for each, and use a high number of randomly generated white noise sequences at these points to infer the required parameter (denoted by $\theta_l$, the index $l$ indicating that the parameter is location-dependent). For all $l$, we can also compute a  set of $K$ good candidate explanatory variables $X_{l1}, \ldots, X_{lK}$, which are preferably fast and straightforward to compute; they can be sky coordinates, the number of observations, the variance of the observation times, or any spectral window feature such as its maximum value in some restricted range of test frequencies, the value at a frequency corresponding to the dominant cadence of the survey, or sums of its highest peaks. Then if a relationship $\hat \theta_l = h_\theta(X_{l1}, \ldots, X_{lk})$ can be established using these simulations, the case-wise estimation during data processing can be replaced by a much cheaper computation: for a new time series during mission, say at location $i$, we compute the necessary features $X_{i1}, \ldots, X_{iK}$ for the given time series, and then use the estimated relation $h_\theta(X_{i1}, \ldots, X_{iK})$ to infer the parameter $\theta$. After obtaining $\hat \theta$, the decision about the significance of an observed periodogram maximum $z_{\rm{obs}}$ can be found from equations \eqref{eq:p_fadm}, \eqref{eq:pgev} or by a simple comparison to the obtained quantile.

\section{Application to the Gaia survey}\label{sec_app}

\subsection{Simulations}\label{subsec_sim}

We simulated 1500 constant photometric time series and 1500 sinusoidally varying photometric time series using AGISLab \citep{holletal12}, both with Gaia-like noise \citep{jordietal10}, at each of 3889 sky positions, with the local Gaia time sampling,  in the following manner:
\begin{enumerate}[1.]
\item We selected several different sets of locations: 
\begin{enumerate}[(i)]
\item 900 locations evenly distributed along an ecliptic meridian between the points $(\lambda = -2^\circ, \beta = 0^\circ)$ and $(\lambda = -2^\circ, \beta = 90^\circ)$; 
\item on a rectangular grid cutting into the most densely sampled ecliptic latitude\footnote{The inclination of the spin axis of the satellite was fixed differently for the mission. As a consequence, the most densely sampled latitude is now at $45^\circ$.} $\beta = 42^\circ$;
\item 714 uniformly randomly distributed points over the sky;
\item a region including part of the Large Magellanic Cloud near the south ecliptic pole, where the Gaia scanning law induces strong aliasing;
\item randomly scattered points over a quarter of the sky with sparse sampling ($N< 55$); and
\item randomly scattered points over a quarter of the sky that had both high aliasing and a low number of points. 
\end{enumerate}
The selected 3889 sky positions are shown in Figure~\ref{fig:map_n_at12} as red dots. 
\item At each location, we simulated 1500 independent white noise sequences and 1500 sinusoidal signals with signal-to-noise ratio SNR $=0.7$ and 1, with uniformly distributed random frequencies in $[0.001,30]\, d^{-1}$, with Gaia-like error distribution, and sampled with the local Gaia nominal scanning law.
\item The full periodogram was computed for every simulated time series (white noise and signals alike), and the maximum of each periodogram extracted.
\item We characterised each location $l$ by the vector of explanatory variables $\{X_{l1}, \ldots, X_{lK}\}$, as described in Section \ref{subsec_parest}. The vector contained the number of observations at $l$, the variance of the observing times, the highest spectral window values $z_4, z_8, \ldots, z_{68}$ in small intervals around 4, 8, $\ldots$, 68 $d^{-1}$ (characteristic Gaia alias frequencies), $S = \sum_{i \in \{4,8,\ldots,68\} } z_i$, 
and various other sums of subsets of these peaks, in an attempt to find the best approximation to an unknown theoretical link between the distribution of  the periodogram maxima and the long-range dependence in the periodogram.
\item At each location, we performed the case-wise estimation of the parameters of the $F^M$, GEV and quantile methods, using the 1500 noise simulations.
\end{enumerate}

We divided the 3889 locations into three parts. The points of the line (region (i) above, see Figure \ref{fig:map_n_at12}) and a band of the densely-sampled rectangle (region (ii) above), in total 1329 points, were selected for a training set. 
The case-wise estimated parameters at the training set locations were used to fit several alternative models for each of the relationships $\hat M = h_{M}(X_{1}, \ldots, X_{k}), \; \hat \xi = h_{\xi}(X_{1}, \ldots, X_{k}), \; \ldots, \; \hat q_{0.99} = h_{q_{0.99}}(X_{1}, \ldots, X_{k})$; these alternative models are described in more details in Section \ref{subsec_modelfitting}. All fitted models were then applied to 1280 locations randomly chosen from the remaining locations (the rest of region (ii) and (iii)-(vi)), and the best one for each of $h_{M},  h_{\xi}, h_{\sigma}, h_{q_{0.95}}$ and $h_{q_{0.99}}$ was chosen according to their performance in reproducing the case-wise estimated parameter values. Finally, the selected models were applied to the 1280 sky positions not used so far, in order to compare their performances on an independent test set.

Since there is no theory predicting the precise form of the functions $h_{\theta}$, we wanted to avoid unnecessary extrapolation as much as possible, so we selected the subsets above such that training, selection and test sets all had sufficient coverage of the whole space of $\{X_{l1}, \ldots, X_{lK}\}$. For instance, the number of observations were in $[45,235]$, $[42,231]$ and $[43,237]$ for the training, model selection and test set respectively, while the sum $S$ was within $[5.7,11]$, $[5.7,10.8]$ and $[5.7,11.1]$ in the three sets.

\subsection{Model fitting} \label{subsec_modelfitting}

\begin{figure*}
\begin{center}
\includegraphics[scale=.73]{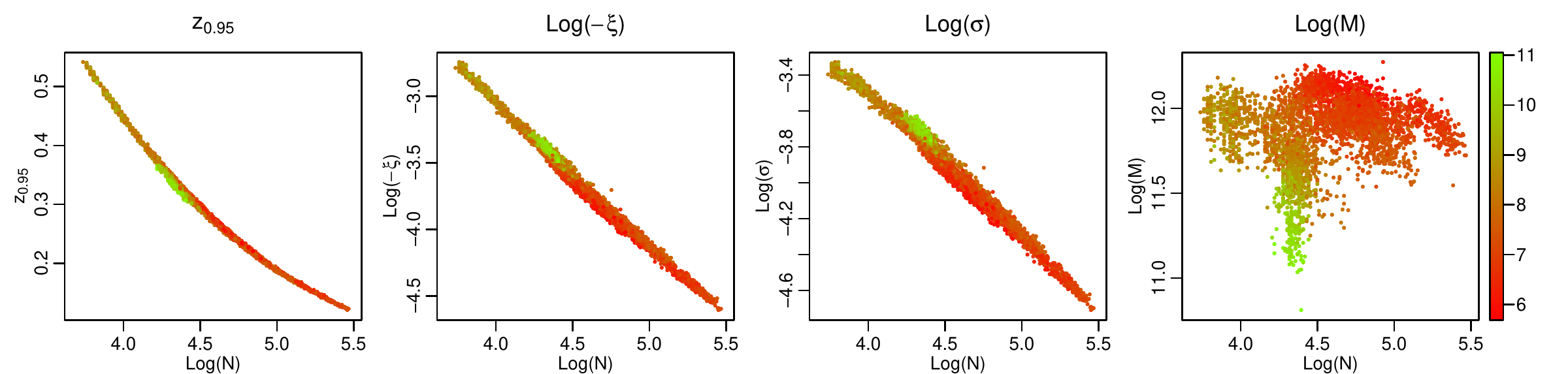}
\caption{Dependences of the parameters of the FAP methods on two important features of the series of observational times. The 0.95 quantile (left panel), $\log (-\xi)$ and $\log \sigma$ parameters of the GEV method (middle two panels) and the logarithm $M$ of the $F^M$ method (right panel) are shown versus $\log N$. The red to green colours code the value of $S$, the level of aliasing for the time series.}
\label{fig:pardep-allmethods}       
\end{center}
\end{figure*}

Figure~\ref{fig:pardep-allmethods} gives a glimpse into the dependence of the parameters of the FAP methods on two of the possible  variables, $N$ and the sum $S = \sum_{i \in \{4,8,\ldots,68\} } z_i$. For $z_{1-\alpha}$, $\xi$ and $\sigma$, there is a well-constrained monotone relationship between the number of observations in the time series and the parameter in question; the link between $M$ and $N$ is rather diffuse, as the dependence on $N$ is at least partly already accounted for by $F$. The relationships for the GEV parameters are not linear, despite the deceptive impression. Moreover, the relationships for all parameters seem to vary somewhat as a function of $S$, indicated on the plots by colour. For $M$, this variation is at least as important as the variation with $N$; for the GEV and the quantile methods, it seems only secondary.   

In the absence of a theory for the relationship between $N$, spectral window summaries and the parameters $z_{1-\alpha}$, $\xi$, $\sigma$ and $M$, several options were tried, ranging from parametric to global non-parametric models:  (a) we classified the spectral windows obtained for 7000 locations randomly scattered on the sky, then fitted separate linear models in each class depending only on $N$; (b) cutpoints in $S$ were used to divide the time series into groups according to the level of aliasing, then separate linear or nonlinear models were fitted within the groups;  (c) we applied nonparametric random forest regression \citep{breiman01} using all the possible covariates we defined for all locations; (d) we fitted 2D thin plate splines using $N$ coupled with a spectral window-related covariate; (e) smoothing splines as a function of $N$, which can be taken as the univariate reduction of the thin plate splines, without a dependence-related covariate. 

These models were fitted using the individually estimated $M$, $\xi$, $\sigma$, $q_{0.95}$ and $q_{0.99}$ at the training set locations. We needed to select both the best type of models and the best set of explanatory variables for all the corresponding links  $\hat M = h_{M},  \hat \xi = h_{\xi}, \hat \sigma = h_{\sigma}, \hat q_{0.95} =  h_{q_{0.95}}$ and $\hat q_{0.99} = h_{q_{0.99}}$, which represents a very broad range of possible models. The random forest regression has some especially useful features: it is able to fit models with a high number of covariates without getting unstable, and it yields a measure of importance about all the covariates, which greatly helps variable selection. Using the few most important variables according to random forest, we fitted the models (b), (c) and (d), together with (a) and (e), and measured the goodness of the models by their predictive mean squared error on the model selection set. Plots of the fitted $\hat M,   \hat \xi,   \hat \sigma, \hat  q_{0.95}$ and $\hat  q_{0.99}$ against the individually estimated parameters were also inspected in order to avoid to select low-scatter, but biased estimators. 

The choice of model variables reflects the lack of theoretical background: it is made purely on the grounds of predictive power and parsimony. Nevertheless, when $S$ was among the best candidate variables, we preferred it to other, more particular choices such as a spectral window value at a specific frequency, since $S$ is a good general summary of the strength of the dependence between two distant frequencies.

For the three method that have parameters to estimate, the best models found are the following.  

\begin{description}
\item{\it $F^M$ method}

\begin{figure}
\begin{center}
\includegraphics[scale=.5]{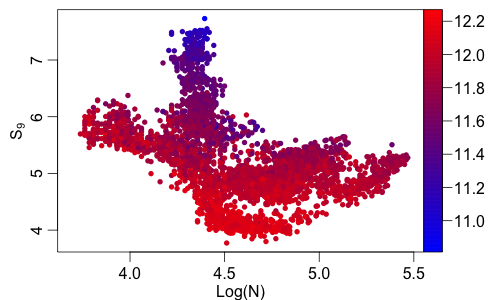}
\caption{The dependence of  $\log M$ on two of the explanatory variables in the best model, $\log N$ and $S_9$. The colours from blue to red indicate the value of $\log M$.}
\label{fig:pardep-m}       
\end{center}
\end{figure}

In agreement with the impressions from the rightmost panel of Figure~\ref{fig:pardep-allmethods}, we selected a relatively complex model with four variables, $N$, $\var(t)$ and two spectral window-related quantities: the sum of spectral window values at nine pre-specified characteristic frequencies, $S_9$, and the sum of the highest three spectral window peaks $S_{\rm{max,}3}$ within the range $[0,70] \; d^{-1}$. There were many nearly equivalent choices, with only marginally worse performance. The final model is taken to be this four-variate random forest model  $\hat M = h_{M}(N, \var(t), S_9, S_{\rm{max,}3}$. Figure~\ref{fig:pardep-m} shows a projection on the 2-dimensional subspace spanned by $S_9$ and $\log N$, with the value of $\log M$ colour-coded.

\item{\it GEV method}

\begin{figure}
\begin{center}
\includegraphics[scale=.5]{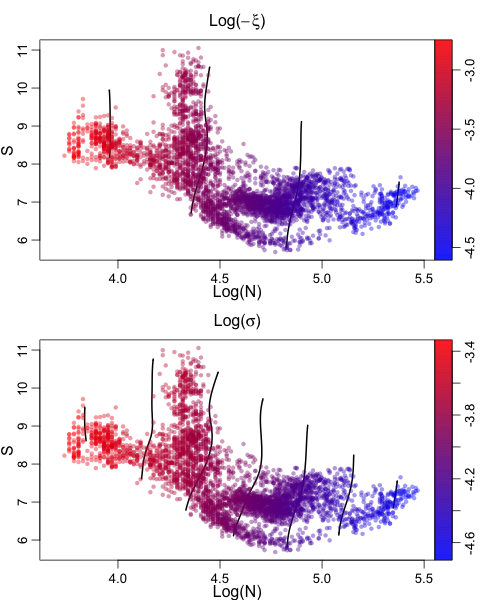}
\caption{The dependence of  $\xi$ and $\sigma$ on the explanatory variables in the best model, $\log N$ and $S$. The colours from blue to red indicate the value of parameters. The 2-dimensional thin plate spline fit is shown as contour lines.}
\label{fig:pardep-gev}       
\end{center}
\end{figure}

For both parameters of the GEV method, the best model in terms of predictive power turned out to be the thin plate spline with covariates $N$ and $S$. An overview of the fit is given in Figure~\ref{fig:pardep-gev}, where the values of the individually estimated parameters are colour-coded over the plane $(\log N, S)$. The fitted thin-plate spline model is superposed as contour lines. Their inclination confirms that to know the parameters of the GEV, the sole knowledge of $N$ would not be sufficient: to estimate $\xi$ and $\sigma$, we also need the value of $S$. Indeed, if we use the smooth spline model without $S$, the fraction of significant p-values on noise sequences becomes increasingly downward biased as aliases grow, and in the highly aliased region around the ecliptic pole, this bias shows a pattern similar to the Baluev method (bottom third panel of Figure~\ref{fig:faprateNbeta}). The fractions become near-independent of region as shown in the bottom second panel of Figure~\ref{fig:faprateNbeta}, once $S$ is included in the fit.

\item{\it Quantile method}

\begin{figure}
\begin{center}
\includegraphics[scale=.5]{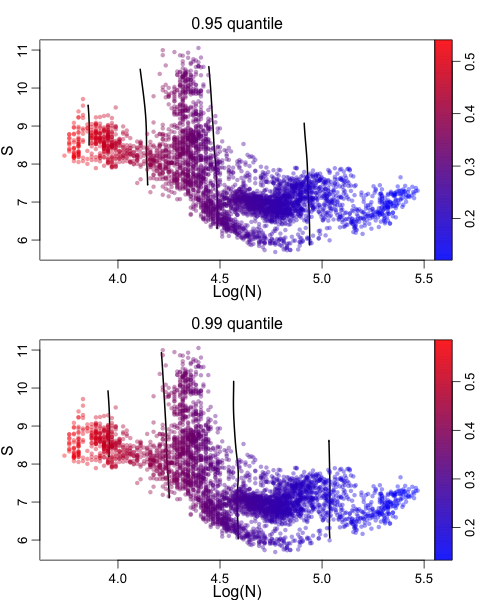}
\caption{The dependence of  $q_{0.95}$ and $q_{0.99}$ on the explanatory variables in the best model, $\log N$ and $S$. The colours from blue to red indicate the value of parameters. The 2-dimensional thin plate spline fit is shown as contour lines.}
\label{fig:pardep-quan}       
\end{center}
\end{figure}

For the two quantiles, we selected the same covariates as for the GEV ($N$ and $S$) among several nearly equivalent candidates. The fitted models are presented in Figure~\ref{fig:pardep-quan}. In this case, the contour lines of the fit seem to be almost parallel to the $S$ axis, implying that we should not find strong effect of aliasing on the results. Despite this, the p-values computed on noise sequences show a bias similar to that of the Baluev method or the smooth spline-modelled GEV: in high-aliased regions, their fraction drops below the nominal levels, though the effect is weaker than for those two models. Consequently, we decided to keep $S$ as a model variable.

\end{description}

\subsection{Quality assessment of the FAP methods}\label{subsec_assess}


We compared the best models selected in Section \ref{subsec_modelfitting} through their statistical size and power, using the noise simulations for the first and noisy signal simulations for the second at the test locations.

\subsubsection{Statistical size}\label{subsubsec_size}

\begin{figure}
\begin{center}
\includegraphics[scale=.8]{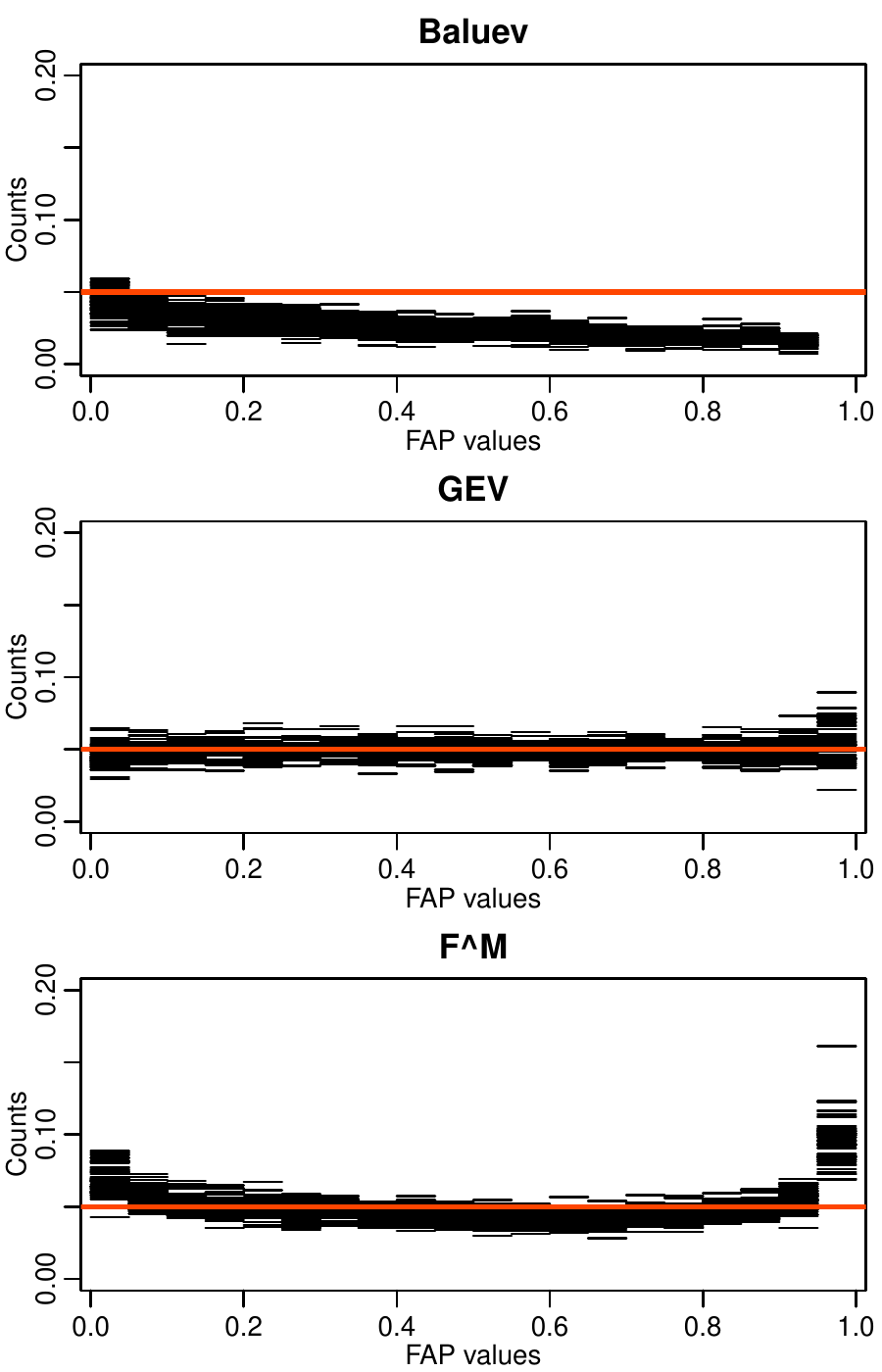}
\caption{Histograms of p-values produced by the Baluev (top), the thin-plate spline GEV (middle) and the random forest $F^M$ (bottom) approximation to the zero distribution of the periodogram peaks. The broken black lines correspond to the histograms of obtained p-values (we omit the vertical lines of the bars for better visibility). The values in the last bin of the Baluev method are scattered between 0.4 and 0.7; they were cut in order to better distinguish details in the lower p-value ranges. The red line represents the expected flat histogram; the more similar the black histograms are to this line, the better the quality of the distributional approximation is.}
\label{fig:faphists}       
\end{center}
\end{figure}

The statistical size is the probability $\alpha$ of a Type I error: that we get a significant test statistic value, although the zero hypothesis is true. In other words, we make a false discovery of a periodicity in noise. The desired value $\alpha$ for a test is always fixed in advance, as the fraction of false detections we allow when applying our tests to white noise sequences. However, it is necessary to make sure that whatever $\alpha$ we may wish to fix in the future, the true proportion of the false discoveries among noise sequences is indeed close to $\alpha$. As was already discussed in Section \ref{subsec_princip}, this hinges on a sufficiently good knowledge of the distribution of the test statistic under the zero hypothesis. If this is unsatisfactory, the p-values computed for any observed periodogram peak $z_{\rm{obs}}$ will be false, and we  either find a lot of contaminating constant stars in the periodic sample, or we lose a higher than expected truly periodic sources in the low SNR regime. 
The first criterion of FAP methods is therefore the quality of their approximation to the true zero distribution of the periodogram peaks. 

To check this, we used the periodogram maxima of the noise sequences simulated at each of the 1280 test set locations. For all of them, we computed the p-values according to each of the models selected in Section \ref{subsec_modelfitting}. The general quality of the approximate distributions was visualised by taking the histograms of these p-values, and comparing them to the uniform distribution. Figure~\ref{fig:faphists} depicts these histograms at 50 randomly selected test locations. 

The overall quality of the approximate distribution is the best for the GEV method, though there are some systematic discrepancies for both very low and very high p-values. The distortion at high p-values, which do not cause any contamination in a p-value-selected periodic object sample, is not interesting from the practical point of view. At the low p-value end, we find slightly fewer values than expected, which means that the approximate distribution somewhat over-estimates the p-values when they are low; this suggests that the FAP based on the GEV will be slightly conservative, tending to give a bit fewer detections than the true distribution would do.

The other two methods have more serious deformations as compared to the uniform distribution. However, the Baluev approximation is not intended to be a true p-value, but an upper bound on its true unknown value. Its discrepancies are concentrated at the uninteresting high p-values, and, agreeing with the findings in \citet{baluev08}, at low p-values it might be used as an approximate p-value. Its performance there is similar to that of the GEV method, somewhat even more conservative. 

The approximation $F^M$ has a systematic curvature throughout the interval $[0,1]$, and underestimates p-values, yielding about 1.5 times more false positives than $\alpha$. The quantile method does not give p-values, so there is no corresponding histogram in Figure~\ref{fig:faphists}.

\begin{figure*}
\begin{center}
\includegraphics[scale=.68]{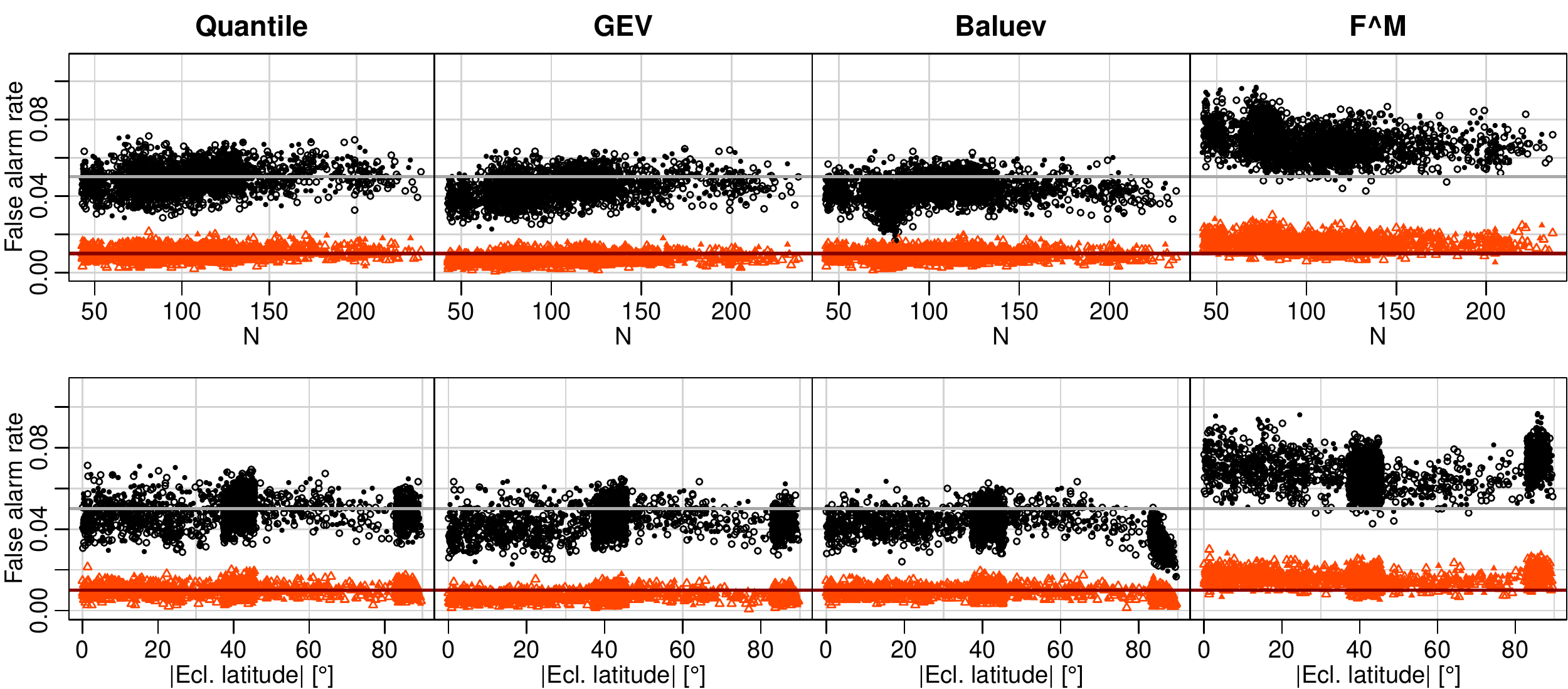}
\caption{False alarm rates of the four methods on 1500 simulated noise sequences at 2248 test locations at confidence level $\alpha = 0.05$ (light grey line) and $\alpha = 0.01$ (dark brown line), as a function of the number of observations $N$ (top row) and the ecliptic latitude (bottom row). The black circles show the fraction of noise sequences found significantly periodic at the level of $\alpha = 0.05$, the red triangles are those significant at $\alpha = 0.01$.}
\label{fig:faprateNbeta}       
\end{center}
\end{figure*}

As there seem to be some, at least slight, discrepancies in the approximate distributions by all methods, it is important to assess how these depend on the features of the time samplings, how they are distributed over the sky, and whether they are capable of causing systematic biases in the detection probability of low-SNR periodic objectsas a function of sky position. For all sky positions, we computed the number of p-values falling below 0.05 (significant at the level $\alpha = 0.05$)  and those below 0.01 (significant at the level $\alpha = 0.01$). Figure~\ref{fig:faprateNbeta} shows these fractions as functions of the number of observations and of the absolute value of the ecliptic latitude. As the 0.95 and the 0.99 quantiles were modelled, the quantile method too could be checked by these means.

One of the covariates in the models is $N$, the number of observations in the time series. If the models found in Section \ref{subsec_modelfitting} are sufficiently good, we do not expect much residual variation of the fraction of significant p-values with respect to $N$. Indeed, this is confirmed in the top row of Figure~\ref{fig:faprateNbeta}, with only slight discrepancies: the GEV method seems to be more conservative than average for low $N$, whereas the Baluev one is more conservative for high $N$. Another set of exceptionally low false alarm rates is present at $N \in [60,90]$ for the Baluev method. For both methods, the effect is weaker for lower $\alpha$. The quality of the modelling depends also on the number of sparsely sampled time series used, so some improvement for the GEV can be expected if we include more locations with sparse sampling into the training set (our present choice overrepresented the densely sampled time series). Unfortunately, no such improvement can be expected for the Baluev method, where the dependence on $N$ is fixed, and there is no free parameter to adapt. The higher false alarm rate of the $F^M$ method, indicated by Figure~\ref{fig:faphists}, is more or less homogeneous over $N$; it is not  obvious whether the impression of higher rates at low $N$ is significant or not. The most homogeneous and best-performing method with the least bias is the quantile method -- perhaps not surprisingly, as it is a direct model for the limit between significant and nonsignificant, and not a model for the whole distribution of periodogram maxima.

Dependence on sky location was not explicitly included in the models, so to check homogeneity with respect to celestial coordinates is important. 
Due to a rough rotational self-similarity of the patterns around the ecliptic axis and the similarity of the two hemispheres (see Figure \ref{fig:map_n_at12}, dependencies on the absolute value of the ecliptic latitude are sufficient for the most important effects. The bottom row of Figure~\ref{fig:faprateNbeta} shows this dependence. Again, the quantile method shows the least systematic variation, followed by the GEV. The $F^M$ method has a weak smooth variation with a minimum false alarm rate around ecliptic latitudes 60$^\circ$. The Baluev method shows a decrease of false alarm rates in the pole regions, due to the fact that the Baluev approximation is built on the assumption of weak aliasing. In the pole regions the time samplings are close to periodic; the spectral windows of these cadences have high peaks at several multiples of the Gaia spin frequency up to high frequencies, as Figure~\ref{fig:spwins} shows. The lower false alarm rates around $N \in [60,90]$, seen in the upper row of plots in Figure~\ref{fig:faprateNbeta}, are due to these locations. Since the formula for the Baluev FAP does not depend on anything else than $N$, further modelling cannot correct for this bias.

\begin{figure*}
\begin{center}
\includegraphics[scale=.8]{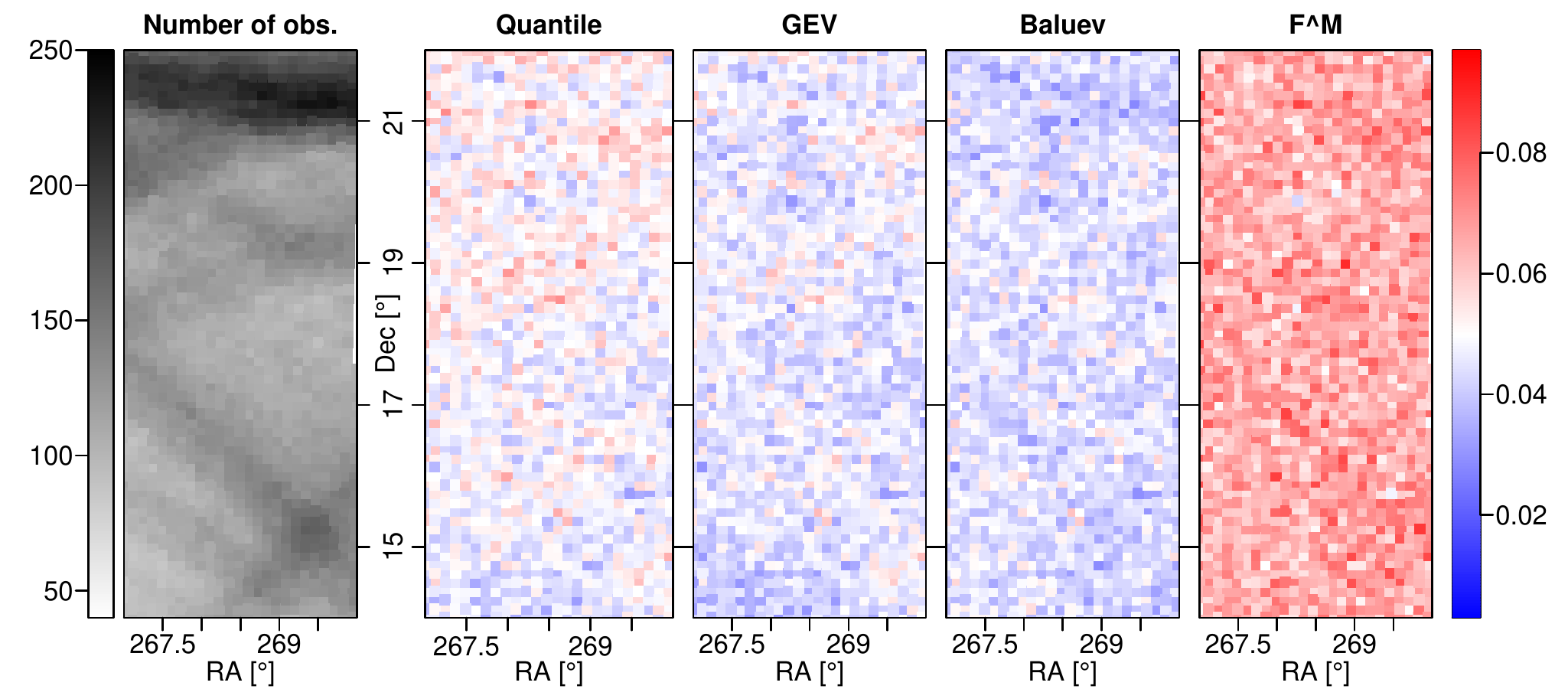}
\caption{Spatial distribution of false alarm rates on the rectangle near the ecliptic equator by the four methods. The grayscale panel on the left shows the number of observations, while the other four panels show the fraction of false positives among noise sequences, using a common colour scale. White corresponds to the nominal $\alpha = 0.05$.}
\label{fig:mapfaprate-ecl}       
\end{center}
\end{figure*}

\begin{figure*}
\begin{center}
\includegraphics[scale=.72]{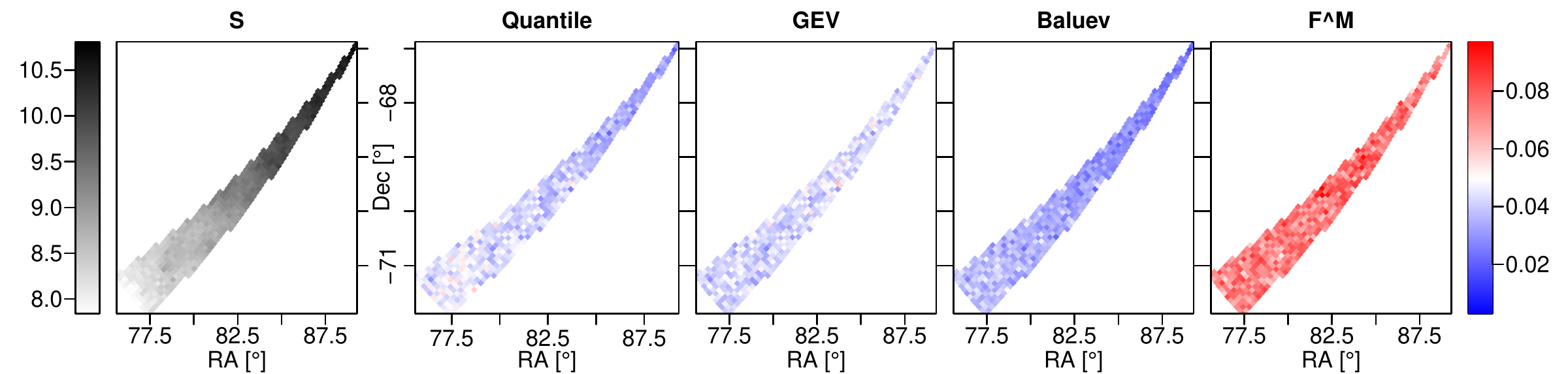}
\caption{Spatial distribution of false alarm rates on the polar region by the four methods. The grayscale panel on the left shows the sum of the dominant spectral window peaks, while the other four panels show the fraction of false positives among noise sequences, using a common colour scale. White corresponds to the nominal $\alpha = 0.05$.}
\label{fig:mapfaprate-lmc}       
\end{center}
\end{figure*}

The spatial patterns of the same false alarm rates in the region of dense sampling (region (ii) in the list of Section \ref{subsec_sim}) are presented in Figure~\ref{fig:mapfaprate-ecl}, for $\alpha = 0.05$. The fraction of false alarms is shown by the colour, white denoting fractions equal to the required level $\alpha$, red indicating too many false alarms, blue too few of them. A homogeneous white colour on the map implies a good-quality method. A map of the number of observations at  the locations is given on the left, in order to compare the patterns of false alarm fractions with those of $N$. The maps confirm the main conclusions from the previous plots, including the (formerly uncertain) impression that the Baluev FAP is increasingly conservative with increasing $N$: the bluest regions coincide with the most densely sampled regions. For the GEV method, there is some tendency of the bluest regions to roughly follow the regions with sparse time sampling, again corroborating the impression of Figure~\ref{fig:faprateNbeta}, though the effect is very weak. The quantile method and the $F^M$ methods show the weakest systematic sky effects, either correlating with $N$, or independent of it.

 The spatial distributions of the false alarm rates in region (iv), at the ecliptic latitude of the Large Magellanic Cloud, are shown in Figure~\ref{fig:mapfaprate-lmc}. The variations in $N$ in this region are less important than in region (ii), but the alias strengths are increasing steadily between ecliptic latitudes 80$^\circ$ and 90$^\circ$, as the leftmost panel shows. Neither the GEV nor the $F^M$ method has discernible correlation with the alias strength, which indicates that most of the variation is accounted for by including dependence-related variables in the fit. The false alarm rates of the Baluev method decrease with increasing alias strengths, as expected from Figure \ref{fig:faprateNbeta}. Surprisingly, the quantile method shows a similar effect, though this was not obvious from Figure~\ref{fig:faprateNbeta}, and the fit for the quantiles also contain a dependence-related covariate. However, the effect is weaker than for the Baluev method, as can be seen from the generally whiter shades in the panel showing the quantile method results than those  in the panel corresponding to the Baluev method.

In summary, the distributional quality seems to be generally best for the GEV method. The Baluev method provides a good approximate distribution in the regime of very low p-values, despite strong overall distortions. Both are slightly conservative. The quality of approximation is the worst for the $F^M$ method, as it has a general tendency to overestimate the significances (underestimate the p-values). Sky systematics are the strongest for the Baluev method, and relatively weak or very weak for the other three.

\subsubsection{Statistical power}\label{subsubsec_power}

\begin{figure*}
\begin{center}
\includegraphics[scale=.75]{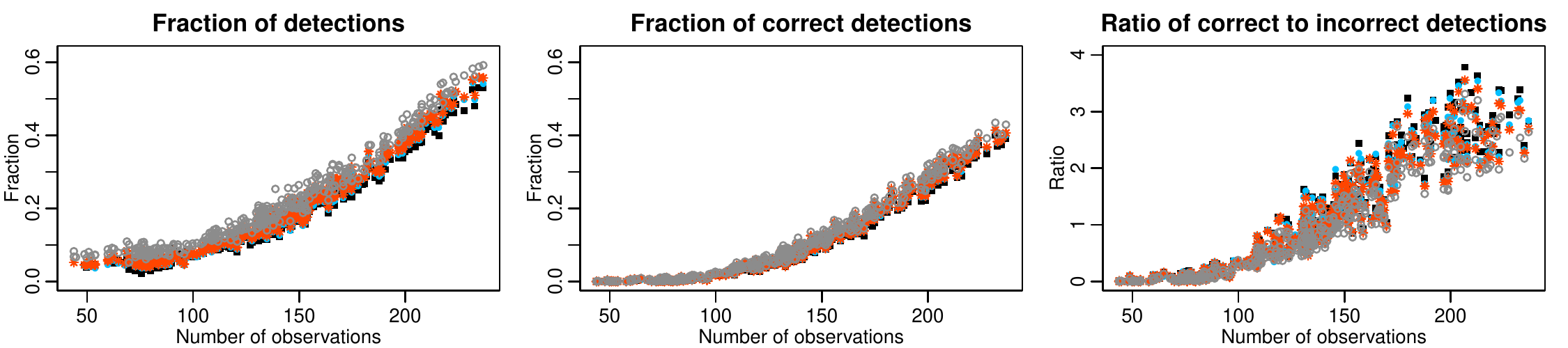}
\caption{Detection and correct detection rates of the four methods on weak signals. Left panel: fraction of all detections on a large number of simulated sinusoidal signals with SNR =1. Middle panel: the fraction of  correct detections among all processed time series. Right panel: the ratio of correct detections to false detections. In all panels, black squares refer to the Baluev method, red stars to the quantile method, blue dots to the GEV method, grey circles to the $F^M$ method.}
\label{fig:fracSignifAmongCorrectFr}       
\end{center}
\end{figure*}

\begin{figure*}
\begin{center}
\includegraphics[scale=.8]{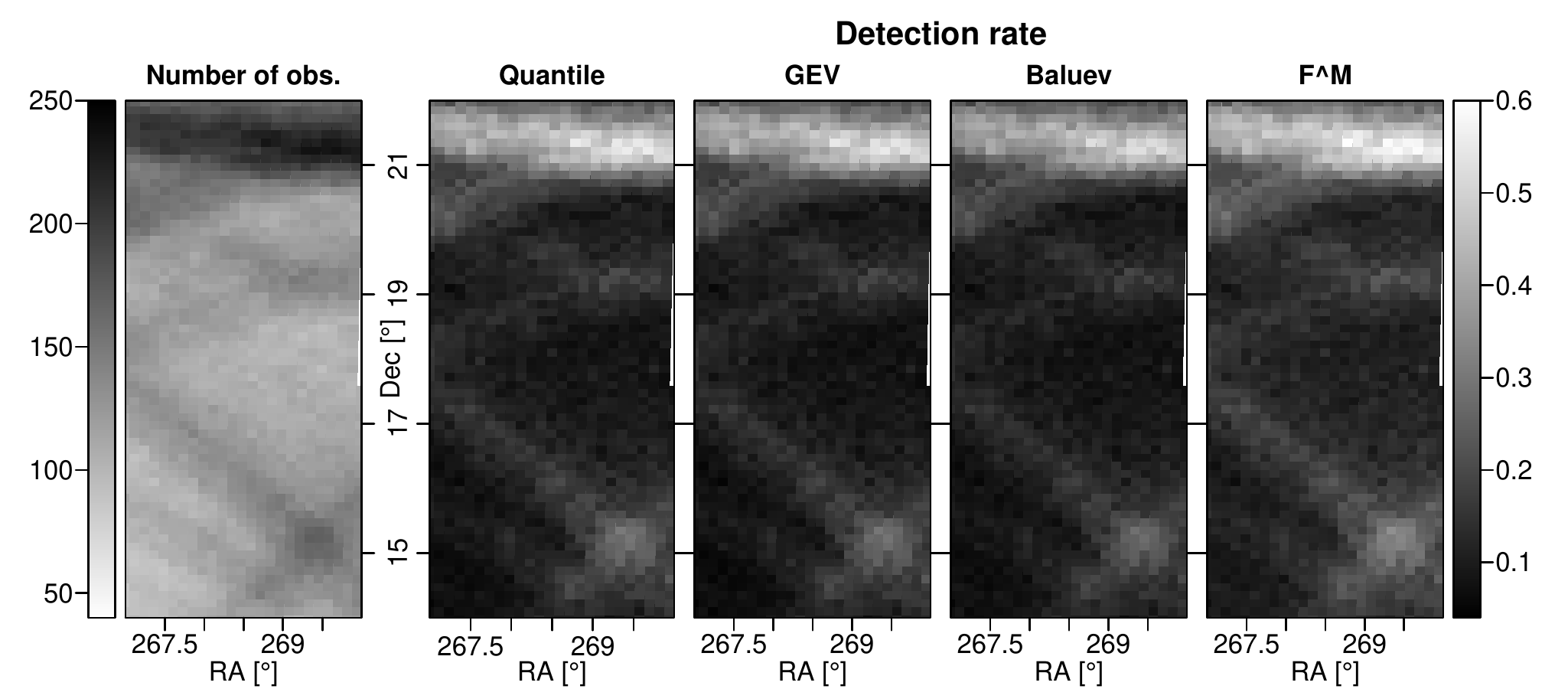}
\includegraphics[scale=.8]{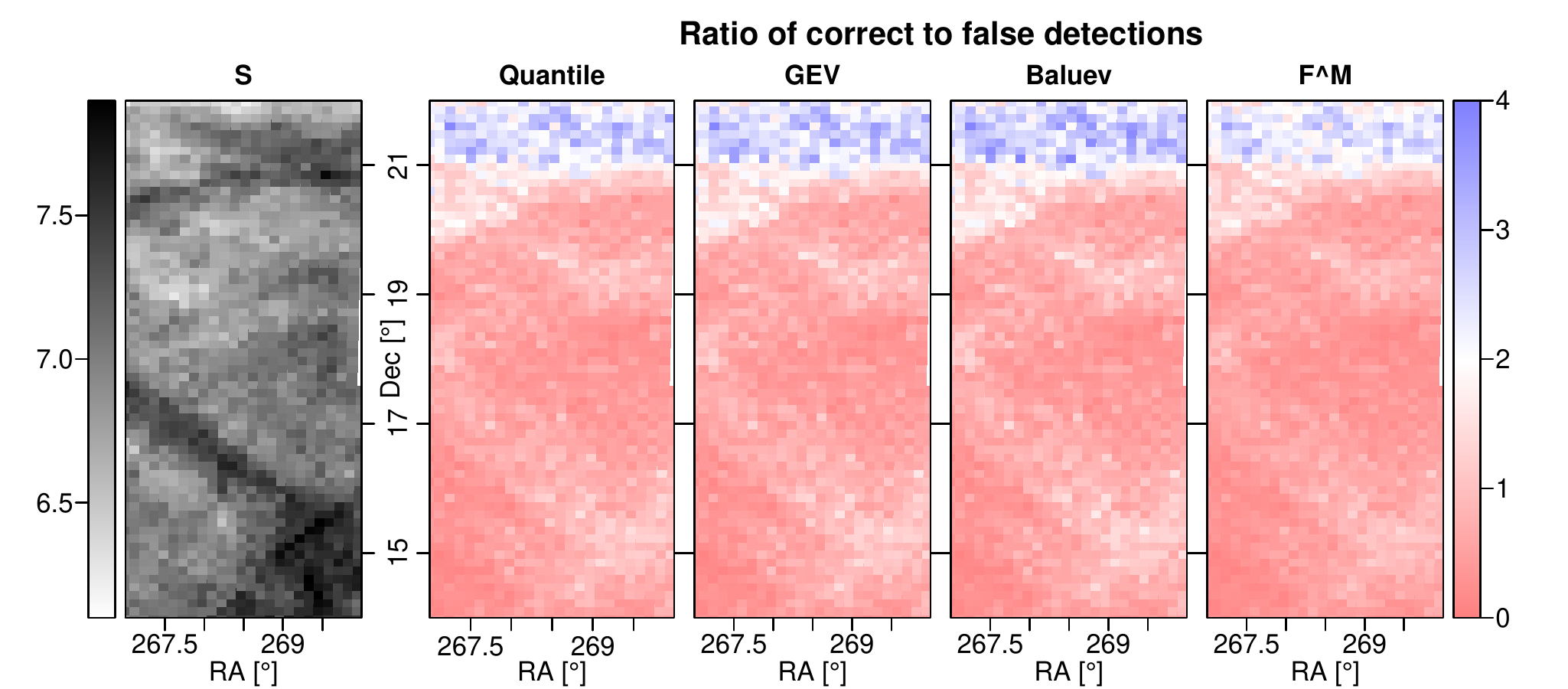}
\caption{Spatial view of the detection rate (top row) and of the ratio of correct detections to incorrect detections (bottom row) on the rectangle at $\lambda=42^\circ$ by the four methods, on 750 simulated sinusoidal signals with SNR =1. The leftmost, grayscale panels show the number of observations (top) and the sum of the dominant spectral window peaks (bottom).  The other four panels in the top row show the fraction of sequences with detected periodicity by the different methods, regardless of correct or false found frequency. In the bottom row, the panels show the ratio using a common colour scale. White corresponds to $r = 2$, blue to ratios higher than this and so  lower contamination, red to lower ratios and thus higher contamination.}
\label{fig:maprecovery10sign05-ecl}       
\end{center}
\end{figure*}

\begin{figure*}
\begin{center}
\includegraphics[scale=.72]{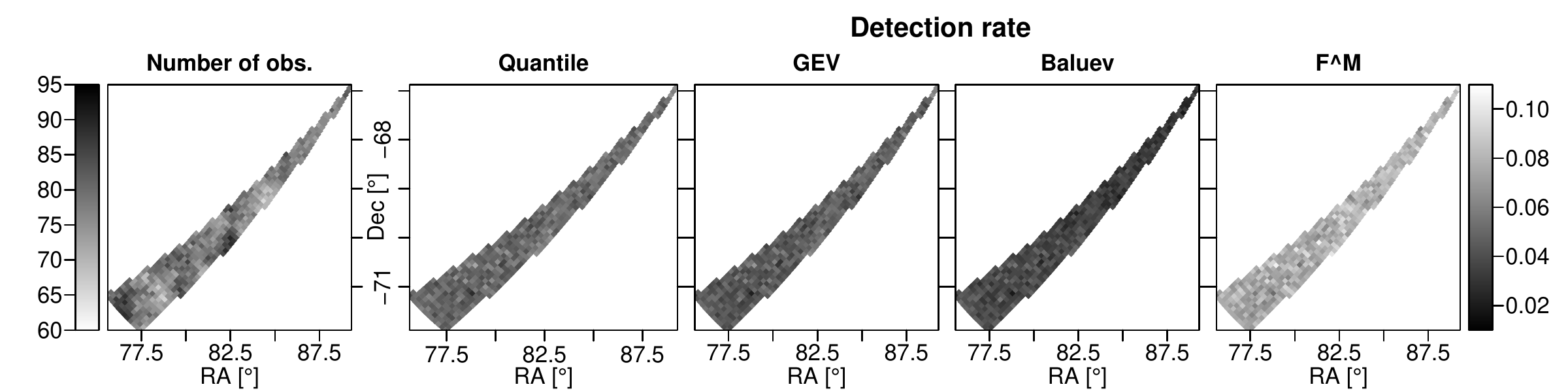}
\includegraphics[scale=.72]{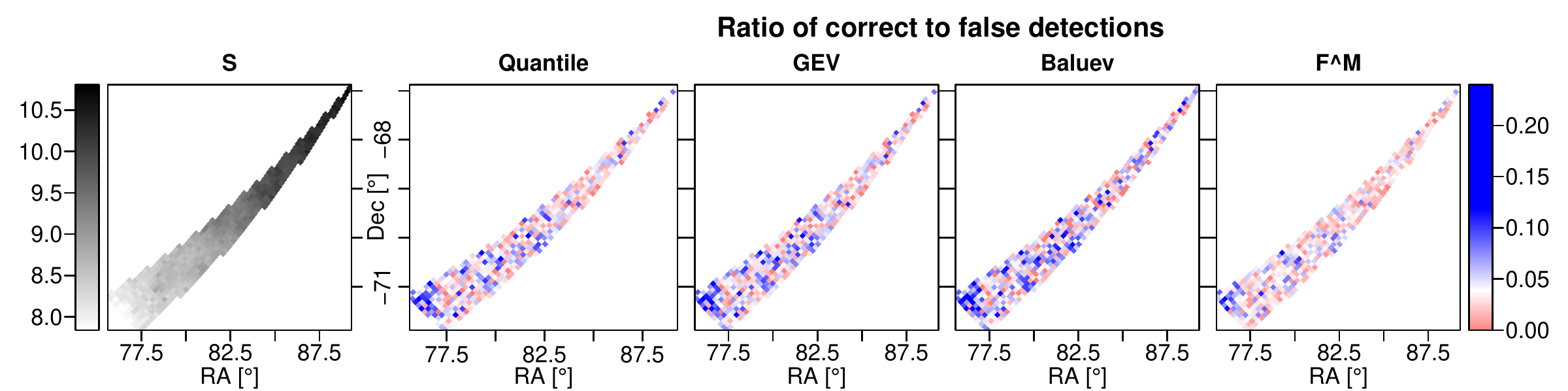}
\caption{Spatial view of the detection rate (top row) and of the ratio of correct detections to incorrect detections (bottom row) in the polar region by the four methods, on 1500 simulated sinusoidal signals with SNR =1. The leftmost, grayscale panels  show the number of observations (top) and the sum of the dominant spectral window peaks (bottom).  The other four panels in the top row show the fraction of sequences with detected periodicity by the different methods, regardless of correct or false found frequency. In the bottom row, they show the ratio using a common colour scale. White corresponds to $r = 0.04$, blue to ratios higher than this and so  lower contamination, red to lower ratios and thus higher contamination.}
\label{fig:maprecovery10sign05-lmc}
\end{center}
\end{figure*}

Statistical power measures the capacity of a method to reject the zero hypothesis when it is not true. 
This quantity is in general hard to compute, most importantly because the alternative hypothesis is usually a composite hypothesis comprising a range of possible alternatives. Simulations for specific cases can be done, in the hope that they yield insight into what can be expected under more involved realistic situations. We used our signal simulations (described in Section \ref{subsec_sim}) at the 1280 test locations. Figure~\ref{fig:fracSignifAmongCorrectFr} shows a general overview of the results for signals with SNR = 1 and using $\alpha = 0.05$. The left panel shows the fraction of all detections, both correct and incorrect, by each of the methods, as a function of the number of observations $N$ in the time series. In agreement with the results on the false alarm rate given in the previous section, the $F^M$ method produces the highest fraction of detection (gray circles), while the Baluev method yields the lowest (black squares).

The middle panel of  Figure~\ref{fig:fracSignifAmongCorrectFr} presents the fraction of correct detections among all the signal sequences. This is, as expected, the highest for the $F^M$ method, and the lowest for the Baluev method, which implies that a sample selected by the $F^M$ method will contain the highest number of correct detections. However,   these correct detections come at the cost of an even higher number of false detections, as the third panel shows: the ratio between correct and incorrect detections is the lowest for the $F^M$ method, and highest for the Baluev method. The difference between the ratios obtained by the four FAP methods is very small for small and near-average $N$, where the detection score is anyway very low for this SNR, but increases at high $N$ where there is a more substantial chance to discover the signal in the noise.

Spatial distributions of the detection rate and the correct/incorrect detection ratio in the densely sampled $\lambda = 42^\circ$ and the highly aliased polar regions are also shown in Figures \ref{fig:maprecovery10sign05-ecl} and \ref{fig:maprecovery10sign05-lmc}, respectively. For the rectangle around $\lambda = 42^\circ$, where the variations in the number of observations are more important than the variations in the level of aliasing, Figure~\ref{fig:maprecovery10sign05-ecl} suggests that the main driving force behind the variations in both the number of detections and the correct/incorrect detection ratio is the number of observations, as the patterns in the four right side panels  follow the patterns of the map of $N$ in the top left panel, rather than the patterns of $S$ in the bottom left panel. Apart from this, the plots confirm both conclusions drawn from Figure~\ref{fig:fracSignifAmongCorrectFr} as to the slightly higher number of detections and the slightly worse correct/incorrect ratio of the $F^M$ method as compared to the other three methods. Those perform quite  similarly, only with tiny differences, which are more visible in Figure~\ref{fig:fracSignifAmongCorrectFr}.   

Similar conclusions can be drawn from Figure~\ref{fig:maprecovery10sign05-lmc} showing the polar region, though with minor differences. With the high $N$ characteristic of this region, the average detection rate is barely above the respective false alarm rates of the methods ($\alpha = 0.05$ was used), and it is found to be quite homogeneous within the region. The ratio between correct and incorrect detections does seem to vary with $N$ and $S$, although it is generally very low, at most 20\% of the detections are correct. The changes with $N$ are obvious for all methods. The variations with $S$ are much less evident; nevertheless, the average ratio seems to decline from $\lambda=-80^\circ$ to the south ecliptic pole, apparently following an average increase of $S$, while the average $N$ is fairly homogeneous throughout the region. 

In summary, whereas the $F^M$ method gives the highest absolute number of detections and of correct detections due to its generally permissive behaviour, its ratio of correct and incorrect detections is the least favourable among the four methods. The Baluev method provides the best correct/incorrect ratio, but in general, there is a small ($\sim$ few percent) loss in the number of sources identified with a correct significant frequency compared to the $F^M$ method. The GEV and the quantile methods perform very similarly to the Baluev one, with a slightly worse correct/incorrect ratio. 

\section{Discussion}\label{sec_disc}

\begin{table*}
\centering
\begin{minipage}{\textwidth}
\begin{tabular}{lllll}
 \toprule
                           & \boldmath${F}^{M}$ {\bf method} & {\bf Quantile method} & {\bf GEV method} & {\bf Baluev method} \\
  \hline
  \hline
   {\bf Supporting theory}    & Independence-based & None & \pbox{3.1cm}{Extreme-value theory: \\ maxima of univariate \\ dependent variables} & \pbox{3.1cm}{Extreme-value theory: \\ upcrossings of stochastic processes}  \\
  \hline
   {\bf Pre-processing}          & \pbox{3.1cm}{Model fitting on a training set of simulated noise \\ Moderate number of simulations required \\} & \pbox{3.1cm}{Model fitting on a training set of simulated noise \\ Moderate--high number of simulations required; \\ number depends on $\alpha$}  & \pbox{3.1cm}{Model fitting on a training set of simulated noise \\ Moderate number of simulations required \\}
   &  \pbox{3.1cm}{Preprocessing not \\ needed} \\
  \hline
   {\bf Distribution}                & \pbox{3.1cm}{Approximate distribution distorted}  & \pbox{3.1cm}{No approximate \\ distribution} & \pbox{3.1cm}{Generally good approximate distribution}   &   \pbox{3.1cm}{Approximate distribution good  at \\ p-values $\leq 0.01$} \\
  \hline
   {\bf Output}                         & p-value & yes/no decision & p-value & p-value \\
  \hline
   {\bf Contamination rate}  &  \pbox{3.1cm}{High (too permissive)} & Low & \pbox{3.1cm}{ Low (slightly conservative)} &  \pbox{3.1cm}{Lowest of all (a bit more conservative than GEV)} \\
  \hline
    {\bf Sky systematics}        & \pbox{3.1cm}{Weak} & Weak & Weak & \pbox{3.1cm}{Moderate} \\
  \hline
    {\bf Robustness}        & \pbox{3.1cm}{Sensitive to distributional discrepancies or outliers} & \pbox{3.1cm}{Insensitive to distributional discrepancies, adaptable to outliers} &  \pbox{3.1cm}{Insensitive to distributional discrepancies, adaptable to outliers}  & \pbox{3.1cm}{Sensitive to distributional discrepancies or outliers} \\
  \bottomrule
\end{tabular}
\caption{\label{table:performances} 
Comparison of the four FAP methods.}
\end{minipage}
\end{table*}

We studied the problem of periodicity detection in large surveys, using the Gaia case to demonstrate the issues and to assess the proposed solutions. Three FAP methods were implemented and tested from the literature: the proposition of \citet{paltani04} and \citet{schwarzenberg-czerny12}, that of \citet{baluev08} and that of \citet{suveges14}, which were all applied to the periodograms from the generalized least squares period search method, one of the most reliable methods used nowadays in astronomy.  We tested also a direct estimation of a critical value between significant and non-significant periodicities corresponding to a fixed confidence level $\alpha$ (the $(1-\alpha)$-quantile).

The quantile method, the $F^M$ method of Paltani--Schwarzenberg-Czerny and the GEV method of S\"uveges involve parameters which depend on the time sampling, and therefore, for the Gaia survey, vary over the sky. Thus, these parameters must be estimated individually for each sky position, or equivalently, for each time sampling. As these functions are strongly varying over short distances, we related the variations in the parameters to the number of observations and the sum of the most characteristic peaks in the local spectral window instead of sky coordinates.  We tested several regression-type procedures to obtain these relations, and selected the best-performing procedure for the parameters of each of the three FAP methods.

The features and typical performances of the FAP methods, each using the parameters estimated from the best-performing procedure, are summarized in Table \ref{table:performances}. The differences between the models are in majority small, apart from one: the generally too permissive behaviour of the $F^M$ method.

The Baluev method provides good FAP estimations in the interesting low p-value regime with some systematics only at the most highly aliased or most densely sampled time series, while not requiring any pre-processing. Thus, it is the least costly but reliable choice for period searches performed on large databases, and can be performed also when a sufficiently good model for the parameters of the other three methods cannot be constructed.

In terms of unbiasedness and sky systematics, the GEV and the quantile methods are the best. However, the quantile method delivers only a decision ``significant/non-significant" referring to a single confidence level fixed in advance, whereas the GEV method produces a p-value which can be compared to any desired $\alpha$, and is the most reliable among the three distributional methods. 
Both the GEV and the quantile methods require a preliminary construction of models for their parameters. Once a model with good predictive qualities is found, the GEV parameters or the quantile can be computed for all time series at small cost. If such a model cannot be found for the typical observational cadences of a database of time series, the parameters must be estimated individually; in that case, the quantile method can require a very high number of full periodogram calculations depending on the desired $\alpha$, whereas the GEV method needs the time equivalent to several ($\sim 10$) periodogram computations \citep{suveges14}.  

The $F^M$ method, in general, underestimates the p-values, and thus, overestimates the significances of the periodogram peaks. However, this effect is quite homogeneous over the sky, and there are only weak systematic location-dependent effects. Like the GEV and the quantile methods, it needs a preliminary model fitting for $M$, and while the models for the first two contain variables that are at least heuristically arguable ($N$ determines the tail of the marginal distribution of the periodogram, while $S$ characterizes the average strength of dependence in the periodogram), the variables necessary for modelling $M$ seem to be more {\it ad hoc}. 
  
The number of detected weak signals is the highest for the $F^M$ method, similarly to the number of weak signals detected with correct frequency, though, due to its too permissive behaviour, the correct detections are hidden in a larger sample of detections. The rate of correct detections among all detections is the best for the Baluev method, and the worst for the $F^M$ method.

The above methods differ from each other in their generalizability to other period search methods and for other surveys than Gaia. The Baluev method is given originally for variants of the least squares method, with three different normalizations \citep{baluev08}. Any periodogram derived from other period search methods should be checked if they have margins compatible with one of these normalizations. The $F^M$ method can be used in every case when the marginal distribution $F$ of the periodogram is known. The GEV method can be used in principle for any periodogram type which indicates the best frequency with a maximum; periodograms which indicate it by their minima can be dealt with by transforming them to be maxima, for example multiplying the periodograms by $-1$. Another advantage is that it is unnecessary to know the marginal distribution $F$ of the periodogram. The predictive regression models must then obviously be set up using the maxima of that specific kind of periodograms. 

As for the extendability of the model building to surveys with different typical time samplings, unfortunately there are no theoretical indications for the direct estimation of the parameters of any of the methods based on the full joint probability distribution of the periodogram. However, there are suggestions from theory that the two most influential factors must be the tail of $F$ (hence the dependence on $N$ for the GLS in this study), and something that characterizes the strength of dependence in the periodogram (hence our choice of spectral window features). Thus, a visualization of individually estimated parameters at a relatively small random set of time samplings of the survey against $N$ and spectral window features can be informative whether such modelling is promising enough.

Our findings, in summary, favour two methods of FAP calculations. One is the GEV method, which yields an overall good performance with weak systematic inhomogeneities over the sky, and can be combined with most period search methods, but needs the setting up of a model for its parameters in a pre-processing step. The other is the Baluev method, which shows some systematic bias over the sky according to the varying strength of aliasing, and can be used only for methods with specific margins, but does not require any pre-processing, and can be computed during processing practically instantaneously. Both methods thus seem to make good progress towards the solution of the question of False Alarm Probabilities in the special case of large surveys.

\section*{Acknowledgments}

\bsp

\label{lastpage}

\end{document}